\documentclass[twocolumn]{aastex631}

\usepackage{gensymb}

\begin{document}

\title{Tracing Red Giant Members of the Globular Cluster Palomar 5 with APOGEE and Gaia}

\newcommand{\affcca}{Center for Computational Astrophysics, Flatiron Institute,162 Fifth Avenue, New York, NY 10010, USA}
\newcommand{\affNBI}{DARK, Niels Bohr Institute, University of Copenhagen, Jagtvej 155A, 2200 Copenhagen N, Denmark}
\newcommand{\afllgcc}{City University of New York (CUNY)--LaGuardia Community College, Department of Natural Sciences, Long Island City, NY 11101, USA}

\newcommand{\RG}[1]{{\color{red} \textbf{#1}}}
\newcommand{\SA}[1]{{\textcolor{purple}{SP: #1}}}

\author[0000-0003-2178-8792]{Allyson~A.~Sheffield}
\affiliation{\afllgcc}

\author[0000-0003-0256-5446]{Sarah~Pearson}
\affiliation{\affNBI}

\author{Lenin~F.~Nolasco}
\affiliation{Department of Physics and Astronomy, Stony Brook University, Stony Brook, NY 11794-3800, USA}

\author[0000-0002-1691-8217]{Rachael L. Beaton}
\affiliation{Space Telescope Science Institute, 3700 San Martin Drive, Baltimore, MD 21218, USA}

\author[0000-0003-0872-7098]{Adrian~M.~Price-Whelan}
\affiliation{\affcca}

\author[0000-0001-6476-0576]{Katia Cunha}
\affiliation{Steward Observatory, University of Arizona, 933 North Cherry Avenue, Tucson, AZ 85721-0065, USA}
\affiliation{Observatório Nacional/MCTIC, R. Gen. José Cristino, 77,  20921-400, Rio de Janeiro, Brazil}

\author[0000-0002-0134-2024]{Verne~V.~Smith}
\affiliation{NSF's NOIRLab, 950 North Cherry Avenue, Tucson, AZ 85719, USA}

\author[0000-0002-0151-5212]{Rafael~Guer\c{c}o}
\affil{Instituto de Astronom\'ia, Universidad Cat\'olica del Norte, Av. Angamos 0610, Antofagasta, Chile}
\affiliation{Observatório Nacional/MCTIC, R. Gen. José Cristino, 77,  20921-400, Rio de Janeiro, Brazil}

\begin{abstract}
The globular cluster Palomar 5 (Pal 5) is in the process of being tidally shredded as it orbits the Milky Way. Its core is currently at a heliocentric distance of $\sim$21 kpc, near apogalacticon ($\sim$18 kpc), and it reaches $\sim$5-7 kpc at perigalacticon. Pal 5's leading and trailing arms stretch over 20$\degree$ on the sky, making them sensitive probes of the Milky Way's mass distribution. In this work, we search for red giant members of Pal 5 using spectroscopic data from APOGEE DR17 and photometric and astrometric data from \textit{Gaia} DR3. Based on position and proper motion, we identify eight members of Pal 5: six in the core and two in the stream. The clustering algorithm HDBSCAN finds these same eight. We then use chemical tagging with APOGEE abundances to search for additional members across five APOGEE fields overlapping Pal 5. While several dozen candidates are identified, most deviate (some significantly) from known kinematic and color-magnitude trends, suggesting that they are less likely to be true members. We estimate the expected number of giants in the APOGEE pointings based on the area and stellar mass of the streams. Given APOGEE's limiting magnitude, we find that few, if any, new giants are expected, especially if the stream is more diffuse at these locations. Our results support the presence of density variations in Pal 5’s tidal streams, consistent with earlier studies attributing such features to baryonic perturbers in the Milky Way, dark matter subhaloes, or interactions with passing globular clusters.
\end{abstract}


\section{Introduction} \label{sec:intro}
Globular clusters (GCs) are ideal objects for studying the structure of the Milky Way. Since Harlow Shapley used RR Lyrae stars in GCs to map the extent and basic spherical nature of the Milky Way's halo \citep{shapley1918}, GCs have remained critical probes for understanding the chemodynamical history of the Milky Way \citep{sz78, zinn93, mackey04,forbes10, massari19}. The work of \citet{sz78} proposed that the Galactic halo was assembled through the accretion of smaller stellar systems over time, rather than forming in a single monolithic collapse. This idea has been reinforced by studies of GC ages, kinematics and metallicities, which suggest that some GCs form in-situ in the Milky Way while others were accreted from dwarf galaxies \citep{forbes10, massari19}. \citet{massari19} classified GCs into distinct origin groups using kinematics and chemical abundances, identifying GCs that are likely accreted versus formed in-situ. 
\citet{belokurov24} further refined this classification, showing that in-situ GCs are found spatially within the central 10 kpc of the disk while accreted GCs have a more extended and isotropic distribution in the halo. They show that accreted and in-situ GCs separate in the [Al/Fe]-[Mg/Fe] plane, where accreted GCs tend to have lower [Al/Fe] at a given [Mg/Fe] than in-situ GCs. With high-precision astrometry from \textit{Gaia}, we can now study the kinematic signatures of these formation scenarios in unprecedented detail. High-resolution spectroscopic surveys such as APOGEE provide chemical abundances that further distinguish in-situ and accreted globular clusters \citep{meszaros20, horta2020}. These combined data sets allow a more detailed investigation of GC formation histories and their role in shaping the Milky Way's structure.   

Palomar 5 (Pal 5) is an exquisite example of a GC in the process of being tidally disrupted. This, in turn, places constraints on the mass distribution of the Milky Way, particularly the dark matter halo \citep{kupper15} and its shape \citep{bovy16}. The core of Pal 5 is located at a Galactocentric distance of 20.6 $\pm$ 0.2 kpc \citep{pw2019}, in the direction of Serpens. Pal 5's extended leading and trailing arms were first identified by \citet{odenkirchen2001}, with SDSS commisioning data. Initially traced over 10$\degree$, these streams were later found to extend over 20$\degree$ \citep{grillmair06}, making them among the longest and most prominent GC streams known. Pal 5's leading and trailing arms formed due to the gradual tidal stripping of stars as it orbits the Milky Way's center. The dissolution of Pal 5 may also be facilitated by interactions with black holes within its core \citep{gieles22,roberts2025}. An intriguing feature of Pal 5's leading arm is that it appears to be truncated at declinations below $-6\degree$ \citep{bernard16}. This truncation may be due to external perturbations that influence the evolution of the cluster. Furthermore, photometric and spectroscopic studies have mapped the leading and trailing arms of Pal 5, revealing density variations and morphological features such as fanning \citep{bonaca2020,kuzma22}.

Pal 5's orbit brings it to $\sim$5-7 kpc of the Galactic center at perigalacticon, and interactions with the Galactic bar could cause stars in the leading arm to fan out and create a decrease in density \citep{erkal17, pearson2017,bonaca2020}. Additionally, simulations by \citet{erkal17} suggest that Pal 5's stellar streams should contain density gaps, possibly caused by interactions with dark matter subhalos. Recent work by \citet{ferrone25} shows that close fly-bys with other GCs could induce stream gaps similar to those expected from dark matter subhalo interactions.

In this work, we present a kinematic and chemical analysis of Pal 5’s stellar streams using spectroscopic data from APOGEE-2 DR17 and photometric and astrometric data from \textit{Gaia} DR3. In Section \ref{sec:data}, we describe the data sample that we constructed from APOGEE-2 and \textit{Gaia} DR3. In Section \ref{sec:chem}, we present results for the chemical abundances for our data. In Section \ref{sec:kin}, we explore the kinematics of our data sample. In Section \ref{sec:disc}, we discuss the results. We provide a summary of this study in Section \ref{sec:summ}.

\section{Data and Methodology} \label{sec:data}
\subsection{APOGEE data}

We primarily use spectroscopic data from APOGEE Data Release 17 \citep{APOGEE:2017, APOGEE:2022} and astrometric data from the \emph{Gaia} Mission data release 3 \citep{Gaia-Collaboration:2016, Gaia-Collaboration:2022}.

The APOGEE-1 and APOGEE-2 surveys are spectroscopic surveys of predominantly Milky Way and Local Group stars using the Sloan Foundation 2.5-m telescope at Apache Point Observatory \citep{Gunn:2006} in the northern hemisphere and the 2.5-m du Pont telescope at Las Campanas Observatory \citep{Bowen:1973} in the southern hemisphere. The APOGEE spectrographs are high-resolution ($R \sim 22,500$; \citealt{Wilson:2019}), infrared ($H$-band) spectrographs, and the primary survey targets were selected with simple color and magnitude cuts \citep[the targeting strategy is described in detail in a series of papers][]{Zasowski:2013, Zasowski:2017, Santana:2021, Beaton:2021}. The spectra were reduced using the APOGEE pipeline \citep{Nidever:2015} and analyzed with the APOGEE Stellar Parameters and Chemical Abundance Pipeline (ASPCAP; \citealt{ASPCAP, Holtzman:2018, Jonsson:2020}) using the APOGEE line list \citep{smith2021,cunha2017,hasselquist2016}. ASPCAP provides estimates of stellar atmospheric parameters and individual element abundances for up to 26 species, including C, N, O, Mg, Al, Si, Fe, depending on the star’s temperature, metallicity, and spectral quality.

Typical APOGEE uncertainties are $\sim$100 K for effective temperature, $\sim$0.1–0.2 dex for surface gravity (log \textit{g}), and $\sim$0.05–0.15 dex in abundance ratios, although this depends on the element, stellar type, the number of spectral lines available for each element and the S/N of the spectra \citep{Jonsson:2020}. Gaia DR3 astrometry has typical uncertainties of $\sim$0.02–0.03 mas yr$^{-1}$ in proper motion for the relatively bright stars in our sample ($G<17$) \citep{lindegren2021}.

\begin{figure*}
    \centering
    \includegraphics[scale=1]{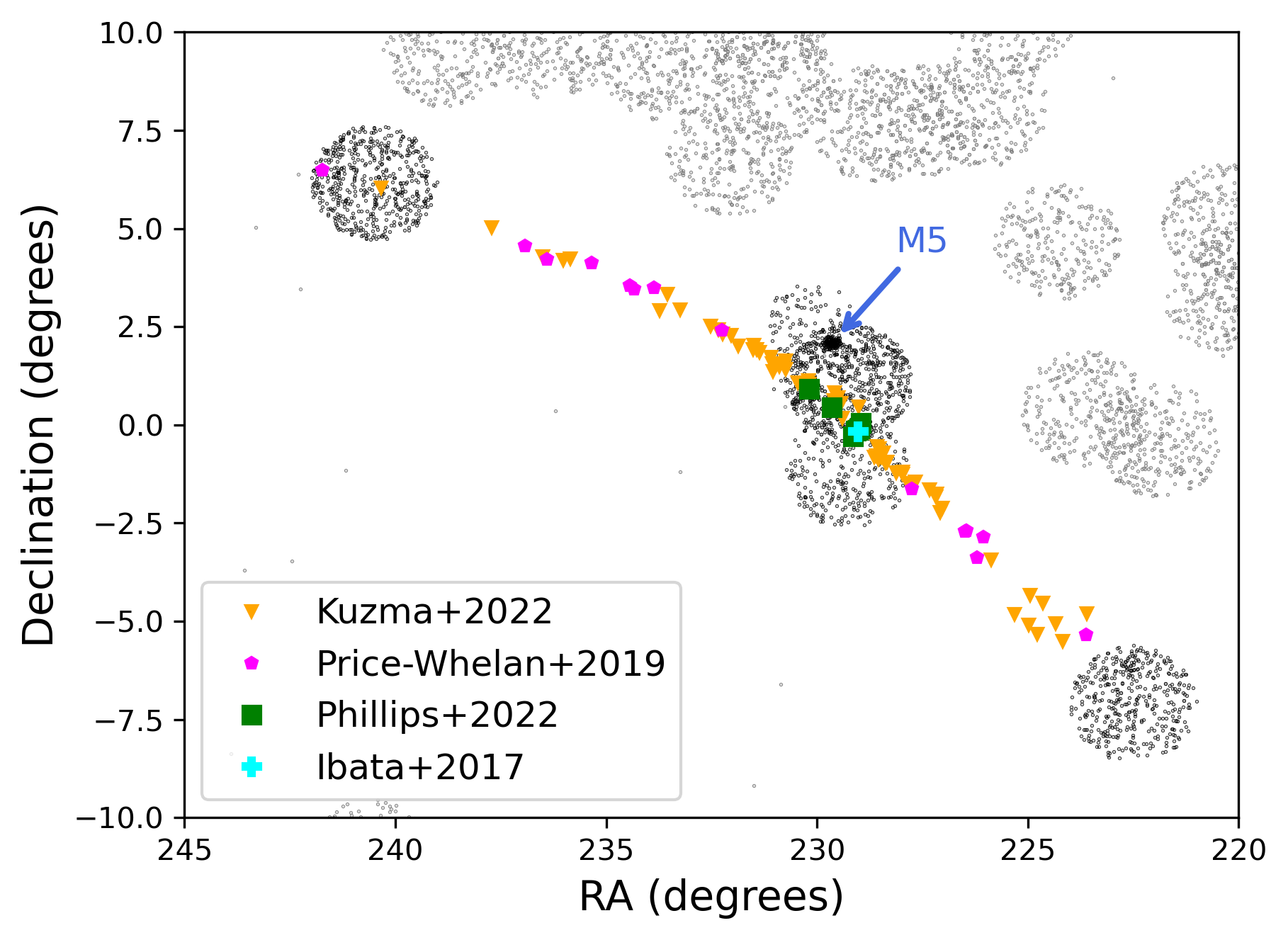}
    \caption{Sky coordinates for the five APOGEE pointings used to identify possible Pal 5 members (black points). The gray points show other pointings nearby, which we did not analyze.  Pal 5 members from the literature are shown as follows: \citet{kuzma22} (orange triangles), \citet{pw2019} (magenta pentagons), \citet{phillips2022} (green squares) and \citet{ibata2017} (cyan plus sign). The cyan plus sign marks the approximate location of Pal 5's core. The small blob of black points at (RA, Dec) = $(230\degree,2.1\degree)$ is the globular cluster M5 and is indicated by the blue arrow.}
    \label{fig:pm}
\end{figure*}

\subsection{Pal 5 Member Selection}
APOGEE dedicated five field pointings to Pal 5 and M5 (M5 happens to lie within 2$\degree$ of Pal 5 on the sky, although it is closer, at a heliocentric distance of 7.5 kpc). Three of these fields surround the cores of Pal 5 and M5, and the other two were observed by APOGEE to sample regions on the sky that overlap with the furthest detections of stars in the leading and trailing arms of Pal 5's tidal streams. We show these five fields as the black points in Figure \ref{fig:pm}.

In Figure \ref{fig:pm}, we also include members of Pal 5 from other studies -- \citet{ibata2017} (cyan plus sign), \citet{pw2019} (magenta pentagons), \citet{phillips2022} (green squares) and \citet{kuzma22} (orange triangles). \citet{ibata2017} carried out two spectroscopic surveys of the Pal 5 core and trailing arm using FLAMES on the VLT and using the AAOmega spectrograph on the Anglo-Australian Telescope. They combined these spectra with deep CFHT photometry to assign membership probabilities and identified 39 likely members of Pal 5, primarily red giants and horizontal branch stars, based on their velocities and positions. \citet{pw2019} combined \textit{Gaia} DR2 proper motions with Pan‑STARRS1 photometry to identify and assign membership probabilities to RR Lyrae stars in the Pal 5 system. They isolated a sample of 27 RRLs (10 in the core and 17 in the streams) using probabilistic modeling in proper motion and distance space. \citet{phillips2022} analyzed seven red giant stars in the Pal 5 system using chemical abundances from APOGEE DR17, cross-matched with \textit{Gaia} EDR3 astrometry. They identified two nitrogen-rich giants located in the trailing arm. \citet{kuzma22} conducted medium-resolution spectroscopy using AAT/AAOmega, targeting red giants in two fields along the leading arm of Pal 5 and identifying 16 new high-probability stream members including eight in an outer field showing evidence of stream “fanning”. These new observations were then combined with previously published RV data (including trailing arm stars) vetted with \textit{Gaia} EDR3 astrometry for a final sample of 109 Pal 5 cluster and stream stars.

To identify members of Pal 5, we first isolate stars in APOGEE that are in the known location of the core, $(\alpha,\delta)$ = $(229.022\degree,-0.112\degree)$, in the window $226\degree < \alpha < 232\degree$ and $-1.5\degree < \delta < 1.5\degree$. To further narrow down the sample, we only select those stars from the position selection that fall inside a proper motion box within $\pm0.2$ mas yr$^{-1}$ of the measured value of ($\mu_{\alpha}^{*}$, $\mu_{\delta}$) = (-2.73, -2.66) mas yr$^{-1}$ from \citet{vasiliev21}. After applying these two selection masks, a sample of eight stars remains. The range in metallicity for these eight stars vary from $-1.33$ to $-1.16$ dex ($-1.23$ $\pm$ 0.05). The metallicities agree well with other values in the literature (e.g., \citealt{smith2002} find [Fe/H] = $-1.3$) and have a small scatter, as expected for a globular cluster. The heliocentric radial velocities range from $-59.7$ to $-56.4$ km s$^{-1}$ ($-57.5$ $\pm$ 1.14) and are also in excellent agreement with previously published values. Of these eight stars, seven were reported as Pal 5 members in \citet{phillips2022} and the other star (2M15160773-0010183) was reported as a member in \citet{ibata2017}. One of the stars (2M15204588+0055032) was observed twice by APOGEE. Six of the eight stars are in the core and two are in the trailing arm, within roughly $1.5\degree$ of the core. 

As an alternative method to identify Pal 5 members, we use the clustering algorithm \textit{Hierarchical Density-Based Spatial Clustering of Applications with Noise} \citep[][HDBSCAN]{campello13hdbscan,loaiza23}. HDBSCAN builds upon the principles of \textit{Density-Based Spatial Clustering of Applications with Noise} \citep[][DBSCAN]{Ester1996} by identifying clusters based on the density of data points. The two main parameters of HDBSCAN are the \textit{minimum number of points} (m$_{\rm Pts}$, from DBSCAN), which determines core and border points, and the \textit{minimum cluster size} (m$_{\rm clSize}$), which defines the smallest number of data points required to form a cluster. We follow the recommendation by \citet{campello13hdbscan} and adopt the same value for both parameters, setting m$_{\rm Pts}$ = m$_{\rm clSize}$ = 2. This choice yielded the lowest standard deviation in metallicity among the clustering outcomes. We find the same eight stars as above, all with 100\% probability of membership, when selecting on proper motion and the abundance ratios [Fe/H], [Mg/Fe], [N/Fe], [Al/Fe] and [Mn/Al]. When selecting only on proper motion, the same eight stars are found, but some of the probabilities of membership are less than 100\%.

\begin{figure*}
    \centering
    \includegraphics[scale=0.6]{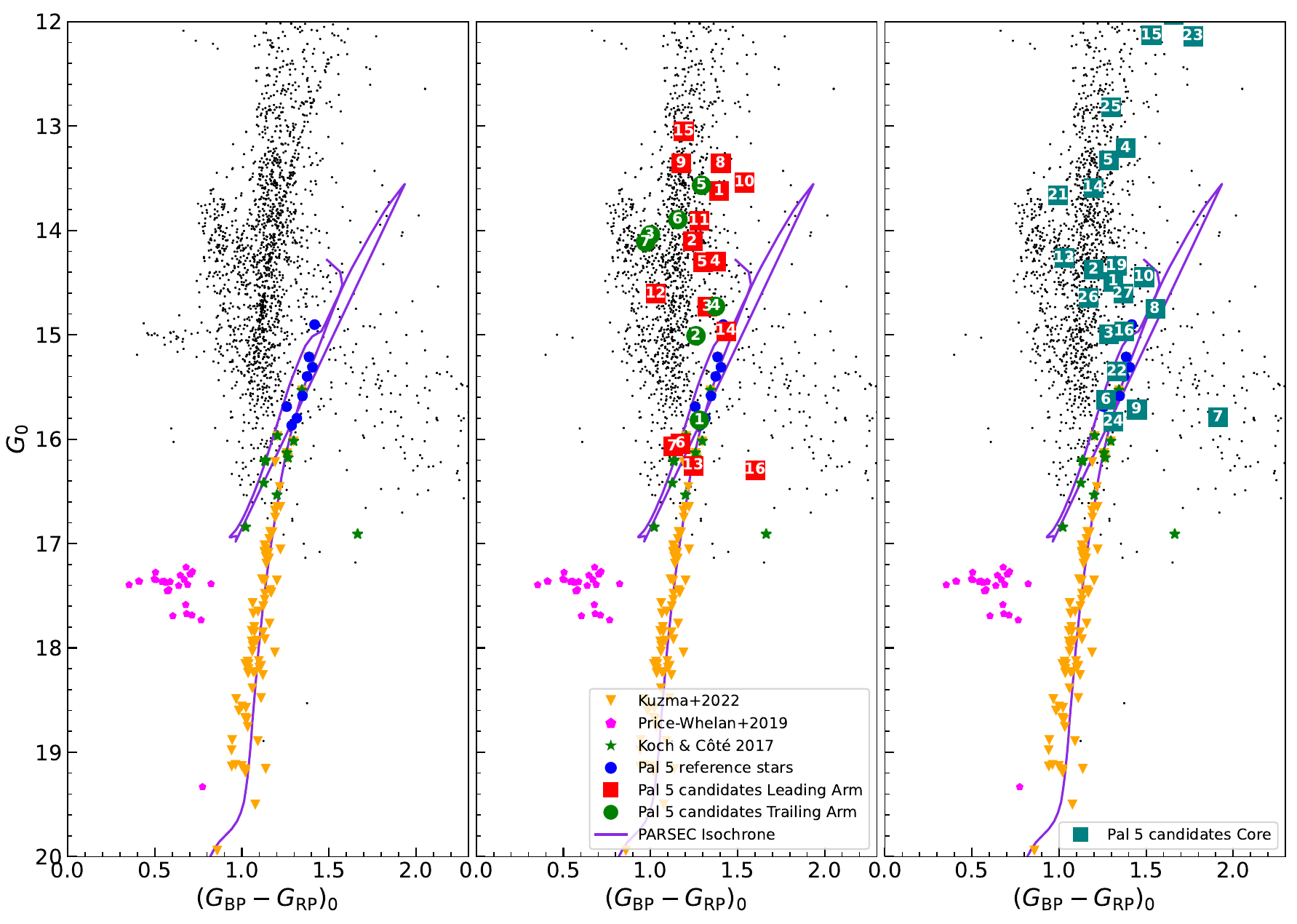}
    \caption{Color-magnitude diagram with \textit{Gaia} photometry. Left Panel: stars in the five APOGEE pointings from Figure \ref{fig:pm} (black points); Pal 5 giants from \citealt{kuzma22} (orange triangles); Pal 5 giants from \citealt{kc2017} (green stars); Pal 5 RR Lyrae stars from \citealt{pw2019} (magenta pentagons); and the eight Pal 5 reference stars (blue circles). 
    An 11.5 Gyr PARSEC isochrone for a population with [Fe/H]=$-1.3$ is shown in purple. Middle panel: The 16 APOGEE stars that were selected chemically for the pointing that overlaps the leading arm, centered at RA = 222.5\degree, are shown as the red numbered squares. The 7 APOGEE stars that were selected chemically for the trailing arm, centered at RA = 240.5\degree, are shown as the green numbered circles. The chemical selection method is discussed in Section \ref{sec:chemtag}. M5 stars have been removed for clarity, using their proper motion signal. Right Panel: CMD for the core candidates, shown as the teal numbered squares.} 
    \label{fig:cmd}
\end{figure*}

We plot the location of these eight Pal 5 reference stars in the color-magnitude diagram (CMD) in the left panel of Figure \ref{fig:cmd} (blue points), with $Gaia$ DR3 photometry. We dereddened the magnitudes using the dust maps of \citet{schlegel98} and the extinction ratios reported in \citet{malhan18}. The left panel of Figure \ref{fig:cmd} also shows the location of Pal 5 members identified in several studies \citep{kc2017,pw2019,kuzma22}. \citet{kc2017} analyzed high-resolution Keck/HIRES spectra of 15 giant stars in the core of Pal 5. The CMD clearly shows the red giant branch and horizontal branch of the globular cluster M5, which lies closer to the Sun and is slightly more metal-rich than Pal 5 \citep{horta2020}. To help isolate the red giants associated with Pal 5, we include an isochrone for an 11.5 Gyr population with metallicity of $-1.3$ dex \citep[PARSEC,][]{bressan12}. We note that the isochrone is not used to determine parameters for Pal 5 but rather to show that members fall along the locus for the same stellar population. We used a distance modulus of 16.6, although since there is a distance gradient along the stream where the leading arm is at closer heliocentric distances \citep[see, e.g.,][]{pearson2017} we expect some scatter above the isochrone for Pal 5 members. For example, the cluster body is at a distance of 20.6 kpc \citep{pw2019}, which equates to a distance modulus of 16.6, while a stream star at a distance of 15 kpc yields a distance modulus of 15.9. The closer stream star would thus have an upward shift of 0.7 mag on the CMD.

\section{Chemical Abundances} \label{sec:chem}
\subsection{Chemistry of GCs}
Globular clusters exhibit distinctive light-element abundance patterns not typically found in field stars. These patterns are the result of early self-enrichment processes within the clusters, in which the first generation of stars polluted the cluster environment with material that later formed a chemically distinct second generation. Second-generation stars often show enhanced N and Al and depleted C and Mg, as well as well-established anti-correlations such as C-N, N-O and Mg-Al \citep[e.g.,][]{bl18, meszaros20}. These abundance anomalies reflect the products of high-temperature proton-capture reactions, specifically the CNO cycle and the NeNa and MgAl chains, which alter elemental abundances through successive nuclear burning steps. Field stars, in contrast, exhibit more diverse chemical signatures reflecting a broader range of star formation and enrichment processes. These chemical patterns are preserved even after stars are stripped from the cluster, making chemical tagging a powerful tool for identifying GC stream members that may be kinematically ambiguous or located in distant, low-density regions of the stream. 

\subsection{Chemical tagging} \label{sec:chemtag}
To chemically isolate Pal 5 stars, we used the abundances of the eight Pal 5 members (``ground truth'') identified via the methodology described in Section \ref{sec:data} as reference points around which to search for stars with similar trends. We exclude stars from M5 by using a proper motion mask, as M5 has a very similar chemical profile as Pal 5 and would contaminate our sample. 

\begin{figure*}
    \centering
    \includegraphics[scale=0.6]{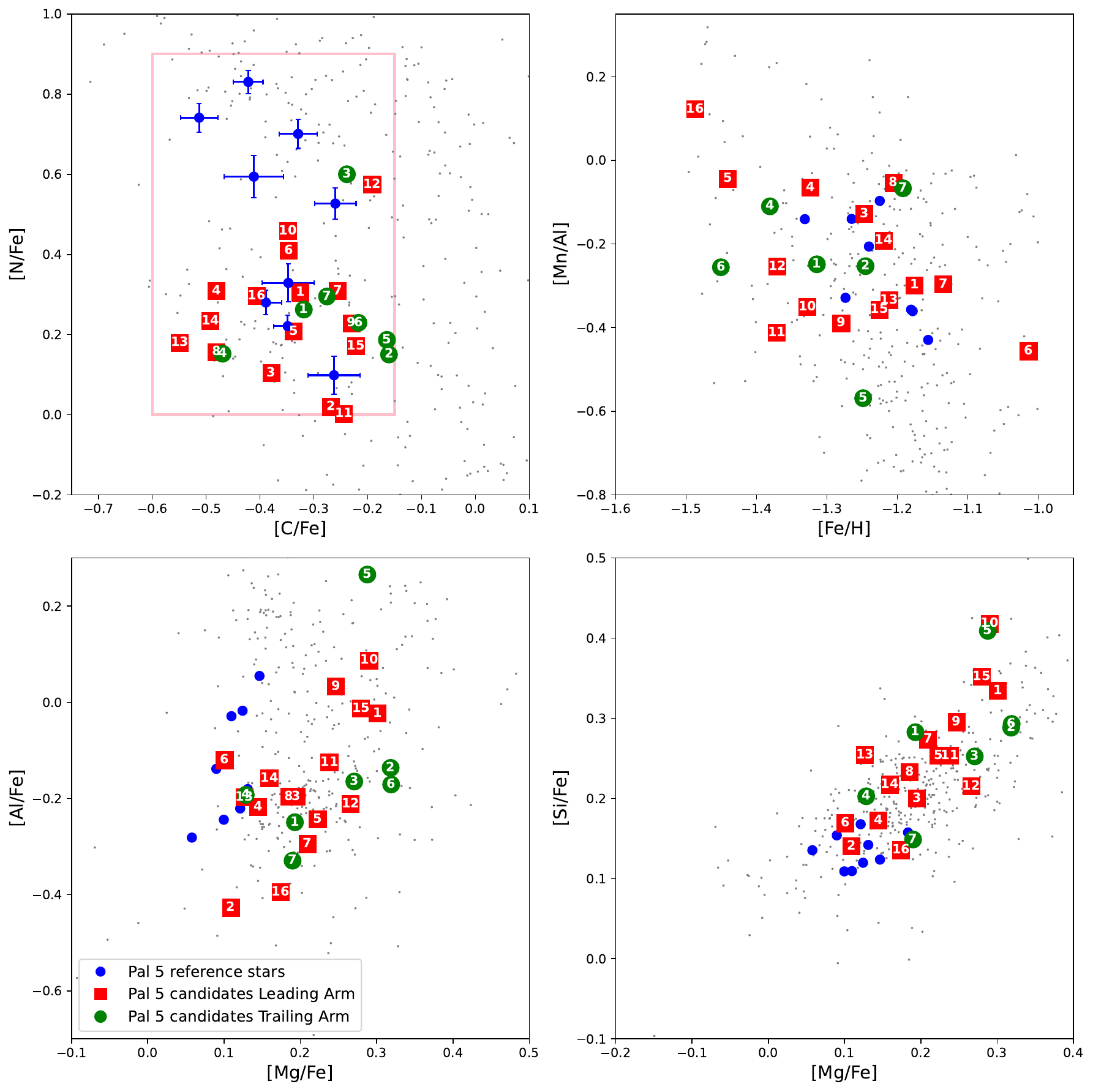}
    \caption{APOGEE DR17 abundances for stars with $-1.5<$[Fe/H]$<-1.0$ in the five APOGEE Pal 5 pointings (black points). The eight Pal 5 reference stars are show as the blue circles (note that there are 9 reference points in blue, as one of the stars was observed twice in APOGEE). We show error bars for these stars in the upper left panel and these error bars are representative of the APOGEE stars in our sample. To be considered a candidate for membership, a star is required to fall within the pink box in the C-N plane (upper left panel) and these are shown as the numbered stars in the red numbered squares (leading arm) and green numbered circles (trailing arm). We show the other three planes for comparison to help in assessing membership.}
    \label{fig:mod_abund}
\end{figure*}

To define the baseline chemical abundance of Pal 5, we show the locations of the eight Pal 5 reference stars in the following planes: [C/Fe] vs. [N/Fe], [Mn/Al] vs. [Fe/H], [Mg/Fe] vs. [Al/Fe] and [Mg/Fe] vs. [Si/Fe], as shown by the blue points in Figure \ref{fig:mod_abund}. We chose these planes because the reference stars form a compact locus in each plane. The [Mn/Al]–[Fe/H] plane separates populations by their nucleosynthetic origins: Mn is produced primarily in Type Ia supernovae, while Al is enhanced in second-generation GC stars. This contrast helps distinguish GC stars from field stars, especially at fixed [Fe/H] \citep[e.g.,][]{hawkins15}. The [Mg/Fe]–[Si/Fe] plane reflects subtle abundance variations that can arise from proton-capture processes at high temperatures, such as those in the MgAl chain. In some GCs, these reactions can convert small amounts of Mg into Si, producing a mild Si enhancement. Although the effect is modest, second-generation GC stars tend to form a tighter sequence in this plane than field stars, making it a useful constraint for chemical tagging \citep[e.g.,][]{horta2020}. We note that Na abundances are not reliably measured in the APOGEE spectra of low metallicity stars given that the Na I lines are very weak.

Next, we only look at stars in the five APOGEE pointings that have $-1.5<$ [Fe/H] $<-1.0$, consistent with a mono-metallic GC population. We note here that we also tried searching for Pal 5 candidates using two additional methods. First, we made selections based on proper motion, radial velocity and metallicity, and then looked at the CMD and chemical abundances to determine membership. Second, we used chemical tagging but with no metallicity cut and the restriction that stars had to fall into all four chemical planes shown in Figure \ref{fig:mod_abund}. These additional searches did not yield any candidates that were consistent in chemistry, kinematics and location on the CMD. We also searched APOGEE fields at declinations lower than $-10$\degree, as model predictions \citep{pearson2017} show that the leading arm could extend to these low declinations.

\begin{figure*}
    \centering
    \includegraphics[scale=0.6]{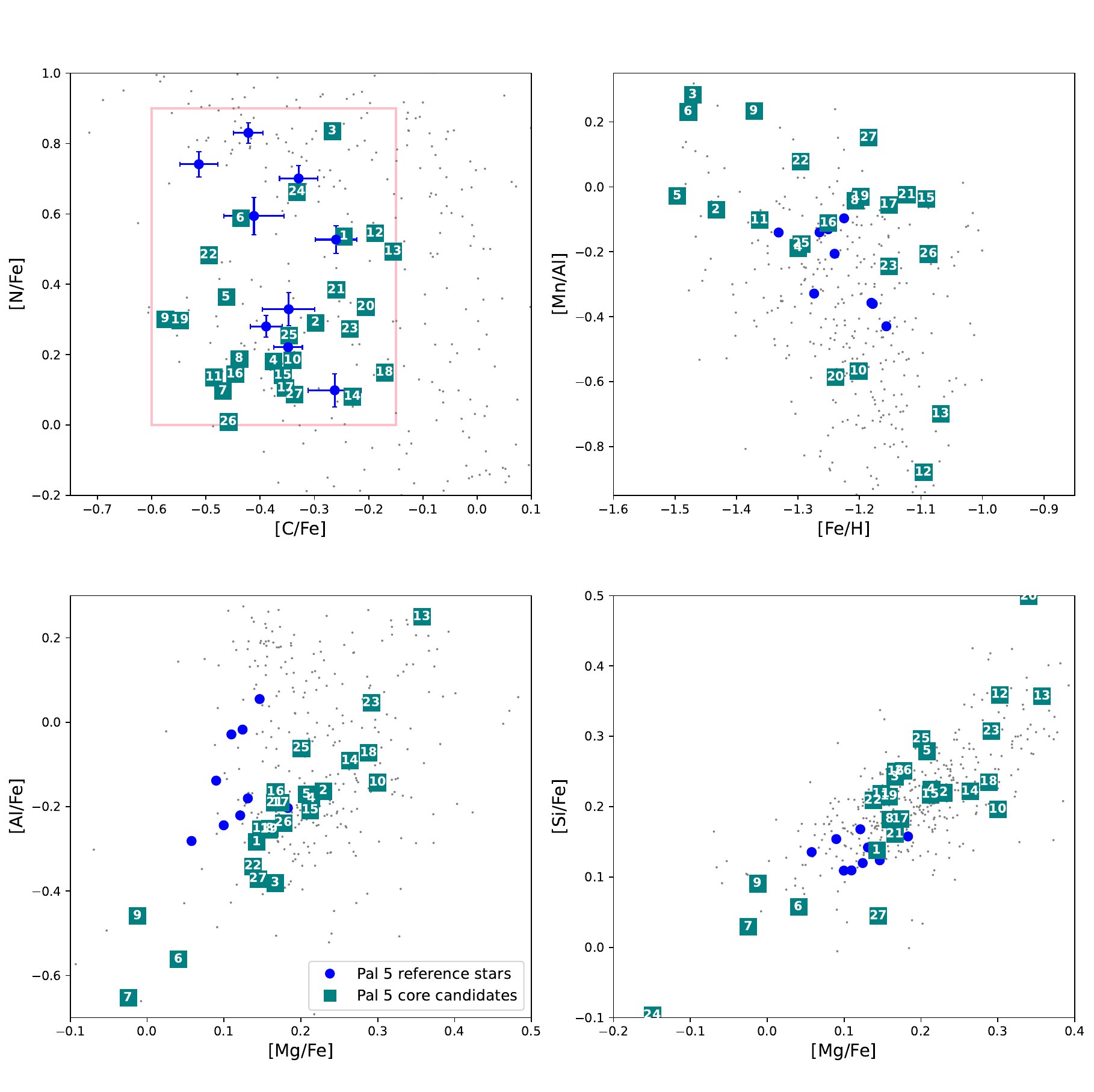}
    \caption{Same as Figure \ref{fig:mod_abund} but now showing Pal 5 candidate members in the core as the teal squares.}
    \label{fig:mod_abund_core}
\end{figure*}

We draw boundaries around the eight Pal 5 reference stars in the [C/Fe]-[N/Fe] plane. These boundaries were selected based on the typical ranges of [C/Fe] and [N/Fe] for GCs from \citet{meszaros20}. If a star falls within the C-N boundary box (pink lines in Figure \ref{fig:mod_abund}), we flag it as a candidate member and check to see where the star falls in the other three chemical planes, as we expect clustering in all of these planes. We do not remove any stars based on their location in the other three planes. We identify several dozen such candidates using this chemical selection method: 26 in the core region (covering the three APOGEE pointings that overlap the core, as seen in Figure \ref{fig:pm}), 16 in the leading arm, and 7 in the trailing arm.
Figure \ref{fig:mod_abund} shows the results for the pointings that cover the leading and trailing arm. The red numbered squares correspond to the leading arm and the green numbered circles correspond to the trailing arm. In Figure \ref{fig:mod_abund_core}, we show the 26 core candidates as the teal numbered squares (note that there are actually 27 stars shown as one of the 26 candidates in the core was observed twice in APOGEE).
We also show these same numbered stars in the middle and right panels of the CMD in Figure \ref{fig:cmd}, with the leading/trailing candidates in the middle panel (same numbering scheme, colors and shapes) and the core candidates in the right panel (teal numbered squares). As seen in Figure \ref{fig:cmd}, only a handful of these candidates fall along the red giant branch and asymptotic giant branch branches for Pal 5. In Table \ref{tab:chem_stars}, we report only the stars that are consistent with the giant branches in the CMD. Although leading arm (red squares) stars $\#3$, $\#4$ and $\#5$ fall above the giant branch, they are included since they are likely closer than 21 kpc, due to their location and the orbital properties of Pal 5. Similarly, a number of the core candidates falling above the isochrone are included in Table \ref{tab:chem_stars}.

\begin{table*}[ht]
\centering
\caption{Chemically and photometrically selected Pal 5 candidate stars in the leading arm, trailing arm, and core.}
\label{tab:chem_stars}
\begin{tabular}{clrrrrrr}
\hline
\textbf{\#} & APOGEE ID & RA (deg) & Dec (deg) & $v_{\mathrm{hel}}$ (km s$^{-1}$) & $\mu_\alpha$ (mas yr$^{-1}$) & $\mu_\delta$ (mas yr$^{-1}$) & {[Fe/H]} \\
\hline\hline
\multicolumn{8}{l}{\textbf{Leading Arm}} \\
\hline
3 & 2M14481587$-$0742550 & 222.06615 & $-7.71529$ & $-68.22 \pm 0.10$ & $-1.07 \pm 0.03$ & $-4.09 \pm 0.03$ & $-1.25$ \\
4 & 2M14483016$-$0655336 & 222.12567 & $-6.92602$ & $311.14 \pm 0.07$ & $-3.35 \pm 0.02$ & $-2.53 \pm 0.02$ & $-1.32$ \\
5 & 2M14490239$-$0730133 & 222.25999 & $-7.50372$ & $-33.86 \pm 0.09$ & $0.37 \pm 0.02$ & $-5.65 \pm 0.02$ & $-1.44$ \\
6 & 2M14493307$-$0813213 & 222.38782 & $-8.22260$ & $-0.96 \pm 0.28$ & $-1.79 \pm 0.06$ & $-1.79 \pm 0.06$ & $-1.01$ \\
7 & 2M14494461$-$0559278 & 222.43591 & $-5.99107$ & $130.74 \pm 0.42$ & $-8.26 \pm 0.06$ & $0.03 \pm 0.06$ & $-1.13$ \\
13 & 2M14531860$-$0715545 & 223.32751 & $-7.26515$ & $-15.65 \pm 0.26$ & $-1.53 \pm 0.07$ & $-1.77 \pm 0.05$ & $-1.21$ \\
14 & 2M14534182$-$0656222 & 223.42429 & $-6.93951$ & $-95.84 \pm 0.08$ & $-1.46 \pm 0.04$ & $-1.61 \pm 0.03$ & $-1.22$ \\
\hline
\multicolumn{8}{l}{\textbf{Trailing Arm}} \\
\hline
1 & 2M15593125+0506267 & 239.88021 & 5.10744 & $-168.66 \pm 0.17$ & $-1.17 \pm 0.04$ & $-1.95 \pm 0.03$ & $-1.31$ \\
2 & 2M15595974+0545526 & 239.99896 & 5.76462 & $-135.26 \pm 0.10$ & $-3.22 \pm 0.03$ & $-4.79 \pm 0.03$ & $-1.25$ \\
4 & 2M16024807+0542234 & 240.70031 & 5.70652 & $-126.48 \pm 0.08$ & $-1.67 \pm 0.03$ & $-2.37 \pm 0.02$ & $-1.38$ \\
\hline
\multicolumn{8}{l}{\textbf{Core}} \\
\hline
1 & 2M15112018+0115351 & 227.83411 & 1.25977 & $102.04 \pm 0.12$ & $-6.01 \pm 0.02$ & $0.58 \pm 0.02$ & $-1.35$ \\
3 & 2M15125817+0117085 & 228.24238 & 1.28571 & $-166.58 \pm 0.16$ & $-0.96 \pm 0.03$ & $-2.72 \pm 0.03$ & $-1.47$ \\
6 & 2M15141111+0030487 & 228.54630 & 0.51355 & $-76.70 \pm 0.24$ & $-2.40 \pm 0.04$ & $-1.79 \pm 0.04$ & $-1.48$ \\
7 & 2M15142413+0143528 & 228.60056 & 1.73135 & $35.08 \pm 0.05$ & $-0.97 \pm 0.05$ & $-0.53 \pm 0.04$ & $-1.22$ \\
9 & 2M15170491+0145472 & 229.27050 & 1.76314 & $-259.23 \pm 0.13$ & $-0.52 \pm 0.04$ & $-0.78 \pm 0.04$ & $-1.37$ \\
10 & 2M15171737$-$0127267 & 229.32238 & $-1.45742$ & $230.26 \pm 0.07$ & $-5.19 \pm 0.02$ & $-1.50 \pm 0.02$ & $-1.20$ \\
16 & 2M15192089+0039294 & 229.83708 & 0.65819 & $-224.99 \pm 0.08$ & $-1.66 \pm 0.03$ & $-2.83 \pm 0.03$ & $-1.25$ \\
22 & 2M15203549+0056416 & 230.14789 & 0.94491 & $-26.31 \pm 0.13$ & $-1.39 \pm 0.04$ & $-1.77 \pm 0.03$ & $-1.30$ \\
24 & 2M15215133+0143131 & 230.46391 & 1.72032 & $55.43 \pm 0.27$ & $-3.19 \pm 0.04$ & $-1.94 \pm 0.04$ & $-1.35$ \\
27 & 2M15240014+0310109 & 231.00060 & 3.16970 & $149.58 \pm 0.09$ & $-3.44 \pm 0.03$ & $-1.32 \pm 0.03$ & $-1.18$ \\
\hline
\end{tabular}
\end{table*}

\begin{figure*}
    \centering
    \includegraphics[scale=0.8]{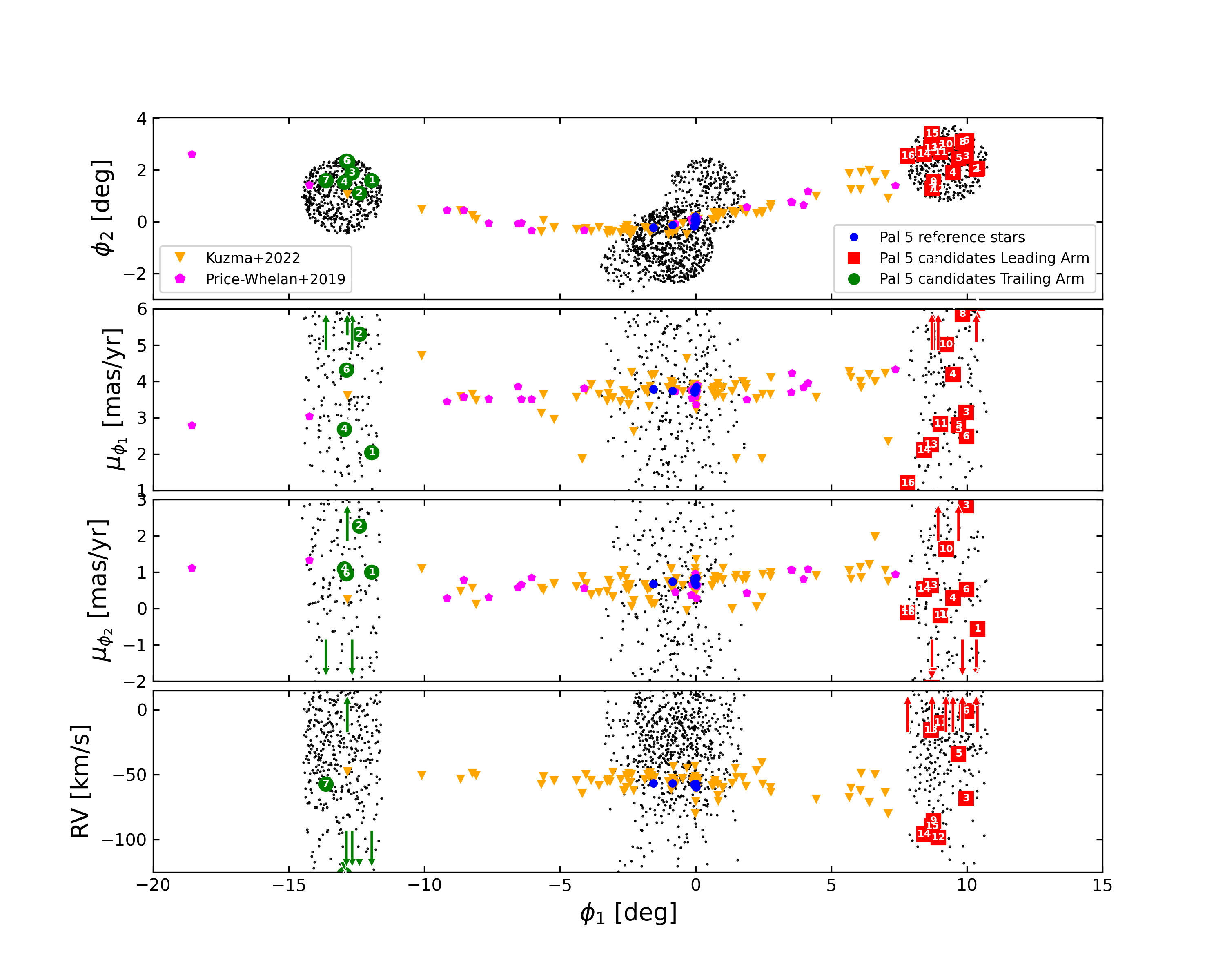}
    \caption{Spatial and kinematic properties of the Pal 5 candidates in the leading arm identified via chemical tagging (see Section \ref{sec:chemtag}). The candidates in the leading arm field are indicated by the red numbered squares and the trailing arm candidates are shown as the green numbered circles. Pal 5 stars from \citet{kuzma22} (orange triangles) and \citet{pw2019} (magenta pentagons) are shown for comparison. Top panel: stream coordinates $\phi_{1}$ and $\phi_{2}$. Second panel: \textit{Gaia} proper motion in $\phi_{1}$ along the stream. Third panel: \textit{Gaia} proper motion in $\phi_{2}$ along the stream. Bottom panel: heliocentric radial velocities from APOGEE along the stream.}
    \label{fig:kin_4pan}
\end{figure*}

\begin{figure*}
    \centering
    \includegraphics[scale=0.8]{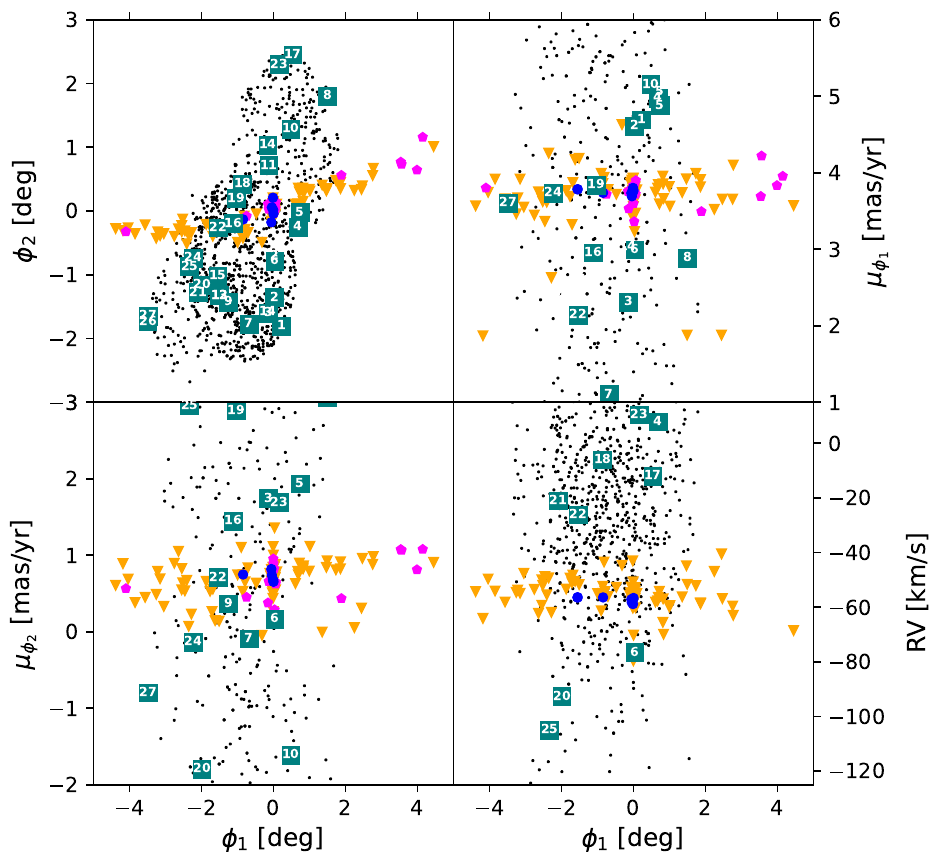}
    \caption{Spatial and kinematic properties of the Pal 5 candidates in the core identified via chemical tagging (see Section \ref{sec:chemtag}). The candidates in the core fields are indicated by the teal numbered squares, and are also shown in the CMD (Figure \ref{fig:cmd}), with the stars from \citet{kuzma22} (orange triangles) and \citet{pw2019} (magenta pentagons) shown for comparison. Top-left panel: stream coordinates $\phi_{1}$ and $\phi_{2}$. Top-right panel: \textit{Gaia} proper motion in $\phi_{1}$ along the stream. Bottom-left panel: Gaia proper motion in $\phi_{2}$ along the stream. Bottom-right panel: heliocentric radial velocities from APOGEE along the stream.}
    \label{fig:kin_4pan_core}
\end{figure*}

\section{Kinematics} \label{sec:kin}
In this Section, we assess the kinematic properties of the Pal 5 candidate members identified through the chemical tagging methodology described in Section \ref{sec:chem}. While their abundance patterns closely resemble those of Pal 5, we use proper motion and radial velocity measurements to determine whether the candidates are dynamically associated with the Pal 5 stream. 

\subsection{Proper Motions}
We adopt the mean value of the proper motion for Pal 5 of ($\mu_{\alpha}^{*}$, $\mu_{\delta}$) = ($-2.73$, $-2.66$) mas yr$^{-1}$ from \citet{vasiliev21}. Stars closer to the cluster or in the inner stream generally follow this motion closely, while stars farther out in the streams are expected to show mild variations, as predicted by stream models such as those in \citet{pearson2017}.

Figure \ref{fig:kin_4pan} summarizes the kinematic properties for stars in the five Pal 5 pointings (black points), with the top panel showing the distribution in stream coordinates $\phi_{1}$ and $\phi_{2}$. The eight Pal 5 reference stars are shown in blue and Pal 5 stars from \citet{kuzma22} and \citet{pw2019} are shown as the orange triangles and magenta pentagons, respectively. The 16 Pal 5 candidates in the leading arm are shown as the red numbered squares and the seven in the trailing arm as the green numbered circles (these numbers are the same as in Figure \ref{fig:mod_abund} and the middle panel of Figure \ref{fig:cmd}). The second and third panels of Figure \ref{fig:kin_4pan} show the \textit{Gaia} proper motions in the stream coordinate system. The trend for Pal 5 in $\mu_{\phi_{1}}$ shows a gentle increase as $\phi_{1}$ increases along the leading arm. This is consistent with the predictions in \citet{pw2019} (see their Figure 2), where $\mu_{\phi_{1}}$ increases at $\phi_{1} > 10\degree$, with the values staying around 4 mas yr$^{-1}$ within the 5-95th percentile confidence interval. A number of candidates in both arms fall into the predicted range, although we note that one of the stars from \citet{kuzma22} falls below the main trend (and we checked the orbit for this star and it does not follow those for the Pal 5 reference stars). For $\mu_{\phi_{2}}$, the values also increase for $\phi_{1} > 10\degree$ towards 2 mas yr$^{-1}$, consistent with the confidence interval from \citet{pw2019}. As with $\mu_{\phi_{1}}$, a number of candidates from both arms that fall along the CMD also follow the general trends for $\mu_{\phi_{2}}$ along the stream.

The proper motion trends along the stream for the core candidates are shown in Figure \ref{fig:kin_4pan_core}. A handful of the core candidates from Table \ref{tab:chem_stars} (e.g., $\#6$, $\#22$ and $\#24$) follow the position and proper motion trends. We next check the radial velocities for all of the candidates in Table \ref{tab:chem_stars}.

\subsection{Radial Velocities}
In addition to proper motions, we examined the radial velocities (RVs) of the candidates in the leading arm, trailing arm, and core. Near the cluster core, the known RV of Pal 5 is approximately $–58$ km s$^{-1}$, as confirmed by multiple spectroscopic studies \citep[e.g.,][]{odenkirchen02, kuzma22}. Simulations of the tidal streams, such as those by \citet{pearson2017}, show that RVs remain roughly constant near the cluster but then increase sharply for stars in the leading arm at R.A. $\approx210^\circ$, which corresponds to $\phi_{1}\approx20\degree$ (see Figure 2 of \citealt{pearson2017}). This steep incline reflects the orbital geometry of the stream and its interaction with the Galactic potential.

For the leading arm candidates (red numbered squares in Figure \ref{fig:kin_4pan}), a few of the candidates from Table \ref{tab:chem_stars} generally follow the main trend with increasing $\phi_{1}$, assuming that the scatter in RV near $\phi_{1} \approx 10\degree$ is real. However, only star $\#7$ from the trailing arm (green numbered circles in Figure \ref{fig:kin_4pan}) falls along the RV trend, and this is not one of the trailing arm stars that follows the CMD trend. Therefore, we do not see any evidence for new members in the trailing arm.

The bottom-right panel of Figure \ref{fig:kin_4pan_core} shows the RVs for the core candidates (teal numbered squares). Of the core candidates that fall along the CMD (Table \ref{tab:chem_stars}), only $\#6$ follows the RV trend (and possibly $\#22$, although it is significantly higher than the main trend).

We checked the expected RVs from simulations of Pal 5, with and without a Galactic bar \citep[][]{pearson2017}. To do this, we re-ran particle spray stream simulations \citep{fardal2015} similar to the simulations in \citet{bonaca2020} using the \texttt{gala} software package \citep{pw2017} for a Pal 5 cluster evolved in a potential including an NFW dark matter halo \citep{navarro1997}, a disk \citep{Miyamoto1975}, and a bar potential \citep{long1992} with a wide range of pattern speeds (20-80 km s$^{-1}$kpc$^{-1}$). We did not find any case where the bar could account for RVs above $-20$ km s$^{-1}$ for stars with RA$>220\degree$. In particular, we found that the Pal 5 stars' radial velocities should shift from $-58$ km s$^{-1}$, turn around at $\mathrm{RA} < 210^\circ$, and then flatten out at around 100 km s$^{-1}$ for declinations lower than $-10\degree$. The proper motions from the models were also within the confidence intervals from \citet{pw2019}.

Any member of Pal 5 is expected to fall along the giant branch on the CMD, have broad agreement with the chemical trends for the cluster, and follow the trends in proper motion and radial velocities. A member could feasibly deviate from the kinematic trends if there is a completely different dynamical process at play which threw the stars out of their orbit. Of all the candidates from Table \ref{tab:chem_stars}, only two remain in the leading arm ($\#3$ and $\#14$) and two in the core ($\#6$ and $\#22$). For a final comparison, we generate the orbits for these four stars and those for the eight Pal 5 reference stars and compare these with the orbits for the eight Pal 5 reference stars using the orbit integrator in the \texttt{gala} dynamics package \citep{pw2017}, all shown in the Appendix (Figures \ref{fig:orbit_pal5ref} -- \ref{fig:orbit6}). None of the orbits for the four candidates match the orbit for Pal 5 members (Figure \ref{fig:orbit_pal5ref}). 

\section{Discussion} \label{sec:disc}
In this study, we searched for red giant members of Pal 5 in two regions along the tidal stream: the leading arm (239\degree $<$ RA $<$ 242\degree, 4.7\degree $<$ Dec $<$ 7.6\degree) and the trailing arm (221\degree $<$ RA $<$ 224\degree, $-5.6$\degree $<$ Dec $<$ $-8.5$\degree), which are both at the limits of previous detections. We also search three regions that overlap Pal 5's core. However, we did not find any red giant members of the Pal 5 in the core or in the leading/trailing streams. Here we discuss whether this is unexpected given the limiting magnitude of APOGEE. 

\subsection{How many APOGEE giants do we expect to find?}
\begin{figure}
    \centering
    \includegraphics[width=0.48\textwidth]{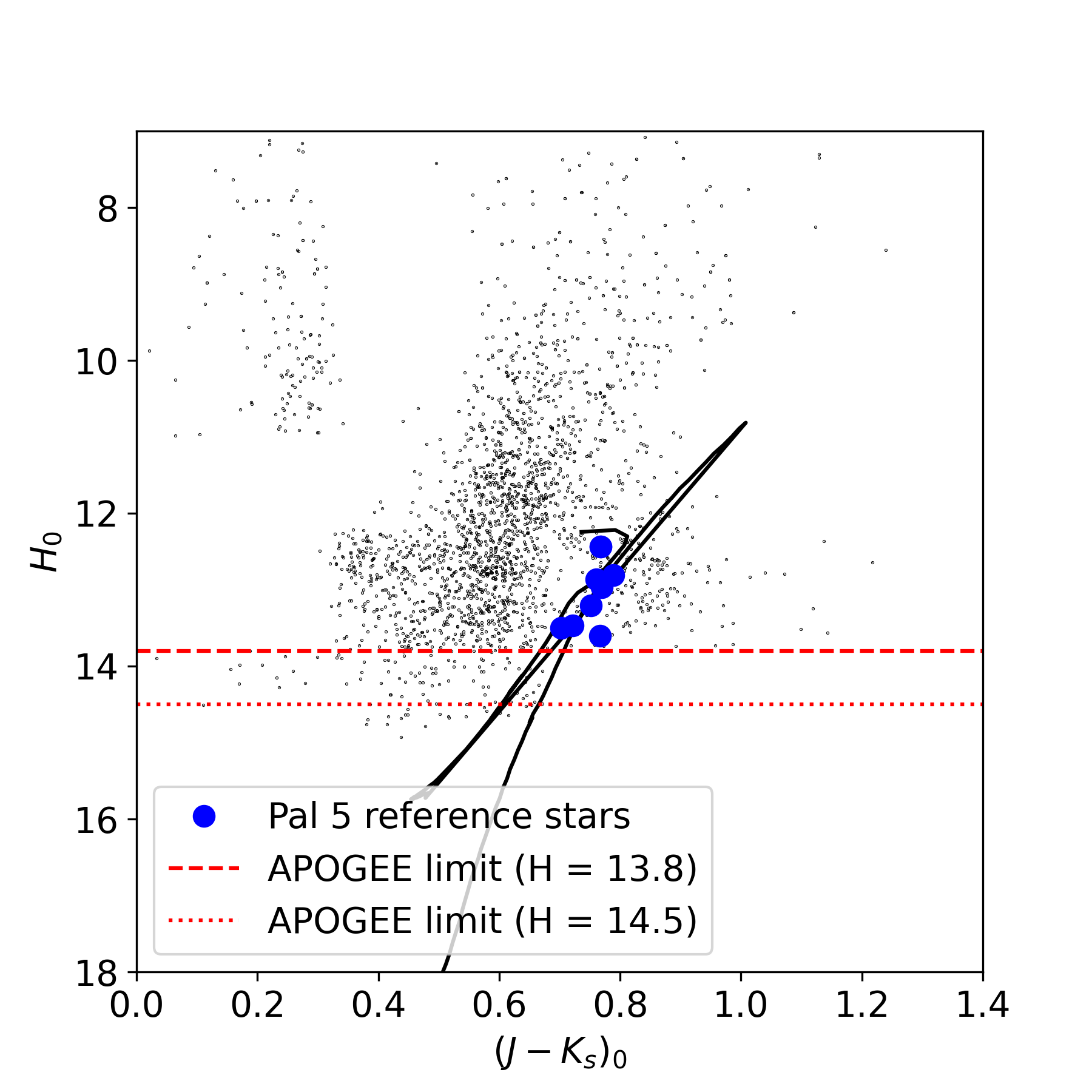}
    \hfill
    \includegraphics[width=0.48\textwidth]{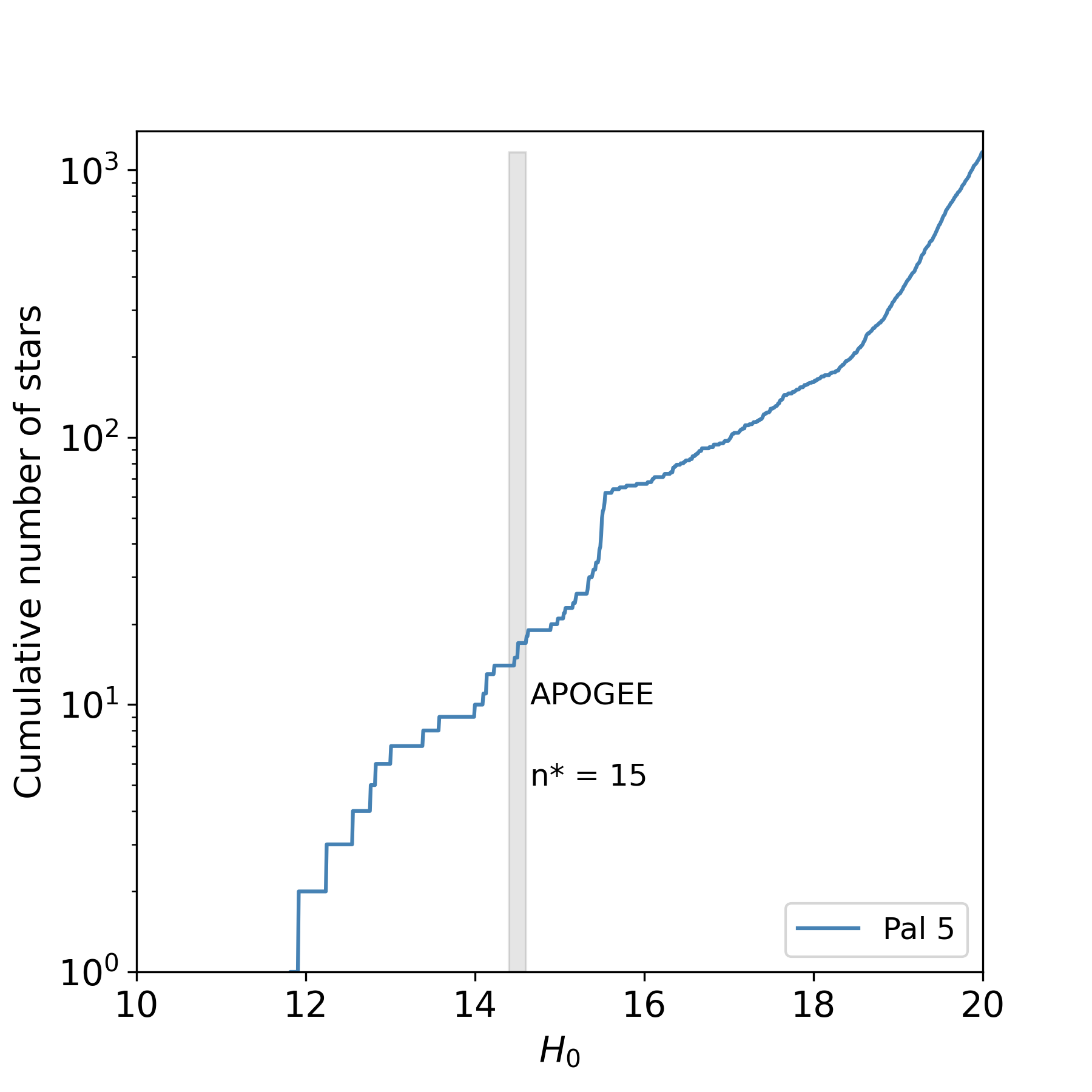}
\caption{\textbf{Left}: 2MASS color-magnitude diagram shifted to the distance of Pal 5. The black points show stars in the five APOGEE pointings explored in this work. The H-band limit of 13.8 for APOGEE is indicated as the red horizontal dashed line and the H-band limit of 14.5 is indicated as the red horizontal dotted line. The PARSEC isochrone is for a stellar population with [Fe/H]=$-1.3$ and an age of 11.5 Gyr. \textbf{Right}: cumulative number of APOGEE giants in the Pal 5 tidal streams we expect to find. For this model run, 15 Pal 5 tidal streams are expected, as noted by the horizontal line at the H-band limit (n* = 15).}
    \label{fig:star_count_3pan}
\end{figure}

To estimate the number of APOGEE giants we expect to find in Pal 5's tidal streams, we use a methodology similar to \citet{pearson19} in their calculation of how many stars from thin streams, like Pal 5, could be detected in external galaxies using deep photometric surveys such as the Roman Space Telescope \citep{spergel2015}. We used a PARSEC isochrone with SDSS and 2MASS bands for an 11.5 Gyr old population with [Fe/H]=$-1.3$ and use a distance modulus of 16.6 to shift from absolute to apparent H-band magnitude. The CMD and isochrone are shown in the left panel of Figure \ref{fig:star_count_3pan}, with lines showing the APOGEE limits of $H_{0}=13.8$ and $H_{0}=14.5$ (some ``special" APOGEE pointings go as deep as $H_{0}=14.5$ and we see some scatter into that region on the CMD). A synthetic stellar population representative of Pal 5 is generated using a Kroupa IMF, and we normalize the synthetic luminosity function by assuming that 3000 stars lie within the apparent magnitude range $20<g_{0}<23$ \citep{pearson19, bonaca2020}. We use the number of stars in the synthetic population that fall within the specified range to scale the synthetic population to match the number of observed stars. We then count how many stars within the scaled population have $H_{0}<14.5$ and find that $\approx$16-24 Pal 5 stream stars are expected to be observed by APOGEE in total. We also count how many stars are expected for $H_{0}<13.8$, which may be more realistic based on the left panel of Figure \ref{fig:star_count_3pan}, and find that $\approx$6-13 stars Pal 5 stream stars are expected. The right panel of Figure \ref{fig:star_count_3pan} shows the cumulative star distribution, with the fainter H-band limit indicated by the gray horizontal line. 
We can get a simple estimate of how many member stars we expect to find in the two APOGEE stream pointings that we analyzed if we assume a linear density and spread the stars evenly across the stream. We do this twice: first, for nine stars, corresponding to the average for the $H_{0}<13.8$ limit, and for 20 stars, corresponding to the average for the $H_{0}<14.5$ limit.

The approximate length of the stream with known detections covers roughly 26\degree, spanning $-18\degree<\phi_{1}<8\degree$; this means that for n* = 9 we expect 1 star roughly every $2.9\degree$ and for n* = 20 we expect 1 star roughly every $1.3\degree$. Each APOGEE pointing is circular with a diameter of $3\degree$ so this should yield roughly 1-2 stars in each pointing for the brighter ($H_{0}=13.8$) limit and 2-3 stars for the fainter ($H_{0}=14.5$) limit. Considering that there are density variations along the stream, these numbers could be even lower. For example, \citet{erkal17} show in their Figures 7 and 9 the expected number counts along the stream for an unperturbed Pal 5 (Figure 7) and Pal 5 perturbed by sub-haloes (Figure 9). In both cases, the linear density along the stream drops significantly at the location of the APOGEE Pal 5 trailing and leading arm pointing locations compared to the regions in the streams closer to the core.
Thus, the fact that we do not find any new giants in APOGEE could simply be due to the limiting magnitude of the survey. 

\citet{bonaca2020} suggested that the leading arm of Pal 5 is dispersed at low declinations. The fact that we do not detect any APOGEE stars in the leading arm could thus be either because the stream
simply does not extend beyond certain declinations or because the stream is ``fanned out'' \citep{pearson2015,bonaca2020}. Previous studies have found that the leading arm has density variations along the stream \citep{erkal17,pearson2017, bonaca2020, kuzma22}. These variations could be due to the influence of the rotating bar on Pal 5 when it is at perigalacticon, which would induce torques on those stream stars and cause them to disperse \citep{pearson2017, erkal17}. While the rotating bar does provide a natural way to disperse stars in a stream, \citet{bonaca2020} found that this alone cannot explain the density variations within the leading arm of Pal 5.

\citet{kuzma22} and \citet{pw2019} detected Pal 5 members in the trailing arm overlapping with the APOGEE pointing (see Figure \ref{fig:pm}). While there are also known density variations in the trailing arm \citep[see e.g.,][]{erkal17}, the reason we do not detect new APOGEE members could simply be due to the limiting magnitude of the survey, as discussed above. 

As a consistency check, we also estimated how many APOGEE giants we expect to find in the core, where six giants have already been observed by APOGEE. 
We use the same synthetic Pal 5 population and find an average stellar mass of $\sim0.83 M_\odot$. We assume two different present day masses for the core: 4,000 M$_\odot$ and 16,000 M$_\odot$, which are similar to the upper and lower limits estimated in \citet{ibata2017} (note that \citet{ibata2017} find a best-fit value of $\approx4300$ M$_{\odot}$).  
We estimate the total present day number of stars in the core by dividing the total mass by the average stellar mass, and we use the same fraction of APOGEE giants as the estimate for the number of APOGEE giants in the Pal 5 tidal streams in the analysis in Figure \ref{fig:star_count_3pan}. For the brighter limit ($H_{0}=13.8$), this analysis for the core yields 5 stars for $M_{core}=4,000$ M$_\odot$ and 15 stars for $M_{core}=16,000$ M$_\odot$. For the fainter limit ($H_{0}=14.5$), this analysis for the core yields eight stars for $M_{\rm core}=4,000$ M$_\odot$ and 33 stars for $M_{\rm core}=16,000$ M$_\odot$. Thus, for the lower mass core this is consistent with the six stars previously detected in APOGEE. 

\section{Conclusions} \label{sec:summ}
In this paper, we analyzed stars in five APOGEE fields to search for Pal 5 members. Three of the fields overlap the core of Pal 5 and the other two fields are located in the leading and trailing arms near the limits of known Pal 5 member detections in the Pal 5 streams. We recover six previously detected stars in the core and two in the trailing arm, within $1.5\degree$ of the core, but do not find any Pal 5 members in either the stream fields or the core. This outcome is consistent with estimates of the number of Pal 5 members expected to be observed by APOGEE, given the survey’s magnitude limits. Our findings support the presence of density variations along the Pal 5 streams, in agreement with previous studies that attribute such structure to interactions with subhaloes, baryonic perturbers, or passing globular clusters. In particular, the lack of detections in the leading arm may reflect either a true truncation of the stream or a fanned, low-density extension that lies below APOGEE’s detection threshold.

\clearpage
\appendix
\section{Additional Orbit Figures} \label{sec:appendix}

\begin{figure}[!ht]
\centering
\includegraphics[width=0.85\textwidth]{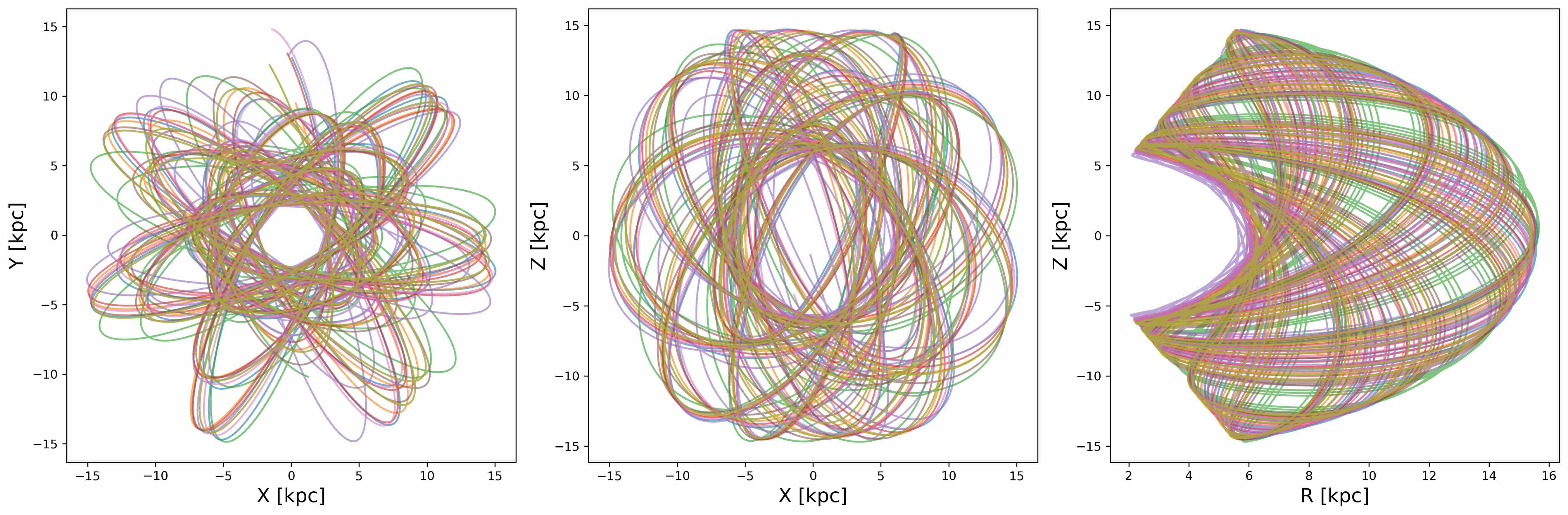}
\caption{Orbits for the eight Pal 5 reference stars. Each color represents a different star.}
\label{fig:orbit_pal5ref}
\end{figure}

\vspace{-0.3cm}

\begin{figure}[!ht]
\centering
\includegraphics[width=0.85\textwidth]{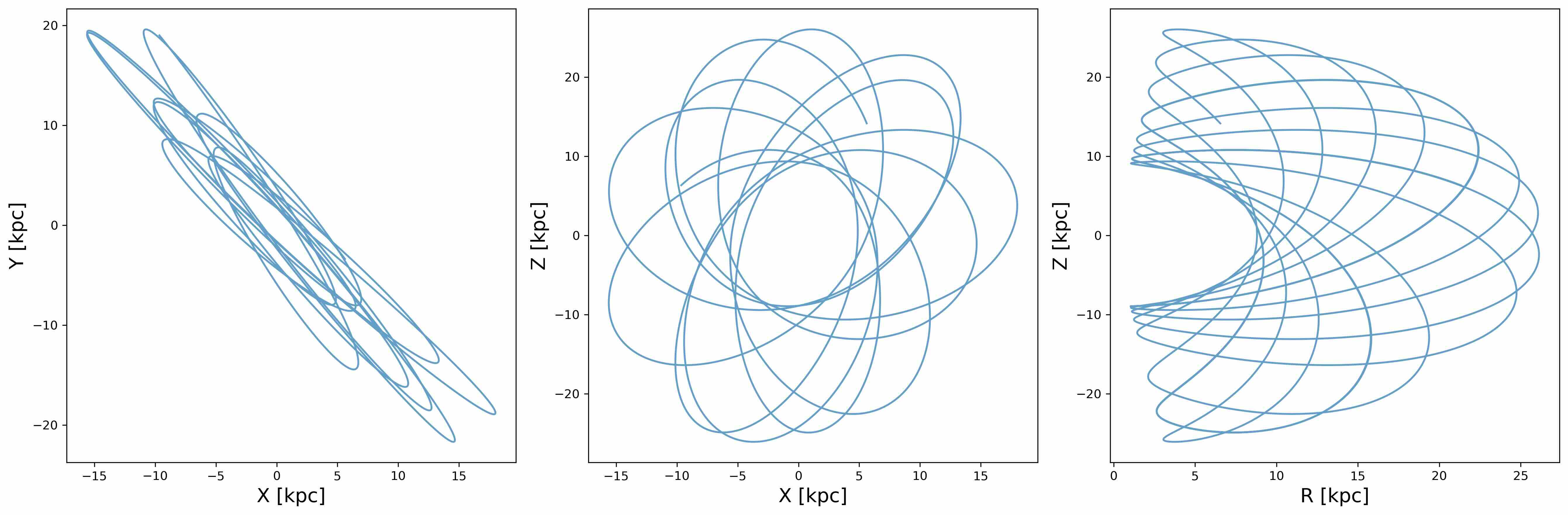}
\caption{Orbit of 2M14481587-0742550 (leading arm candidate 3).}
\label{fig:orbit2}
\end{figure}

\vspace{-0.3cm}

\begin{figure}[!ht]
\centering
\includegraphics[width=0.85\textwidth]{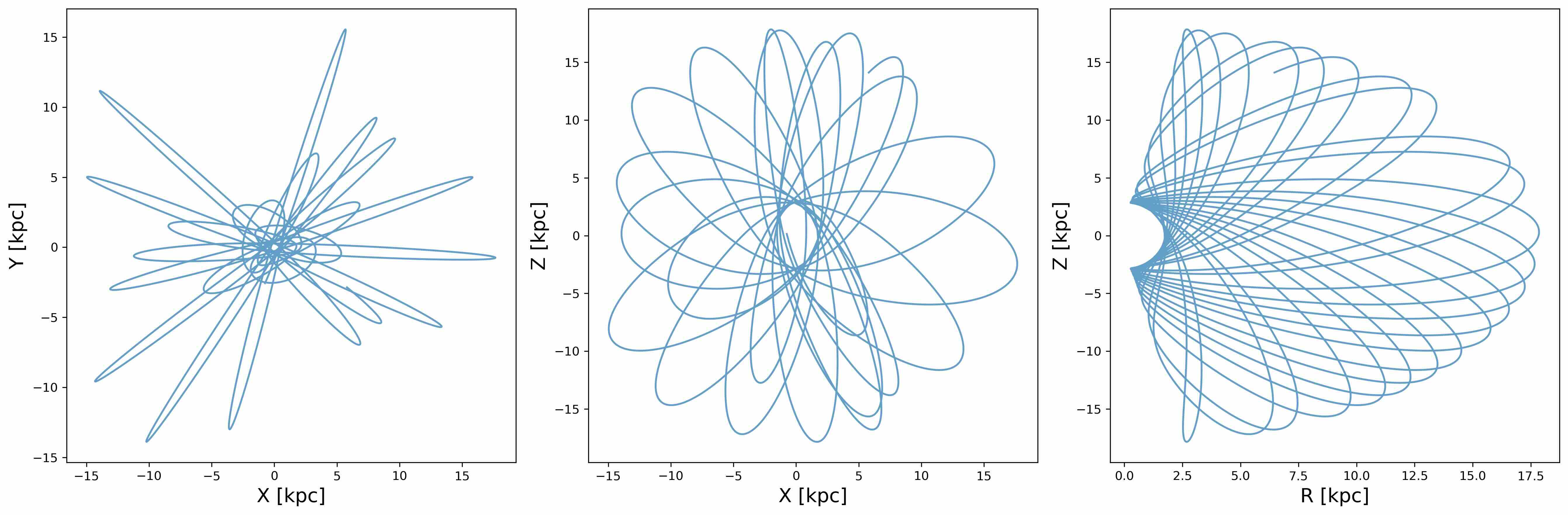}
\caption{Orbit of 2M14534182$-$0656222 (leading arm candidate 14).}
\label{fig:orbit4}
\end{figure}

\vspace{-0.3cm}

\begin{figure}[!ht]
\centering
\includegraphics[width=0.85\textwidth]{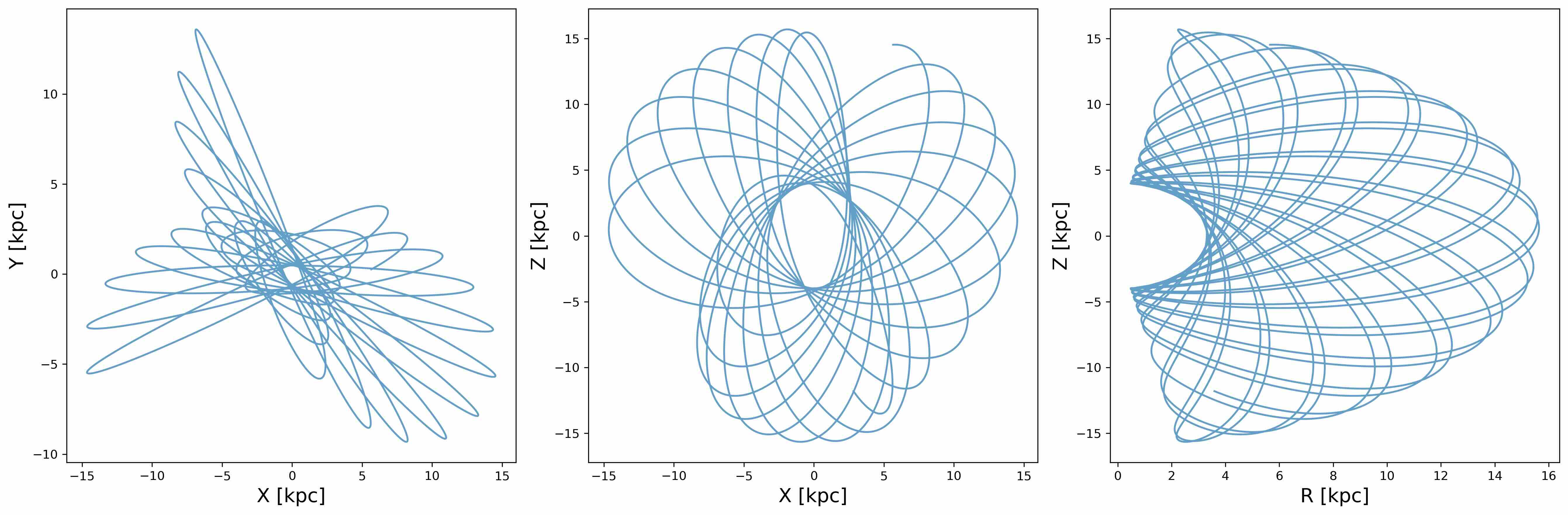}
\caption{Orbit of 2M15141111+0030487 (core candidate 6).}
\label{fig:orbit5}
\end{figure}

\vspace{-0.3cm}

\begin{figure}[!ht]
\centering
\includegraphics[width=0.85\textwidth]{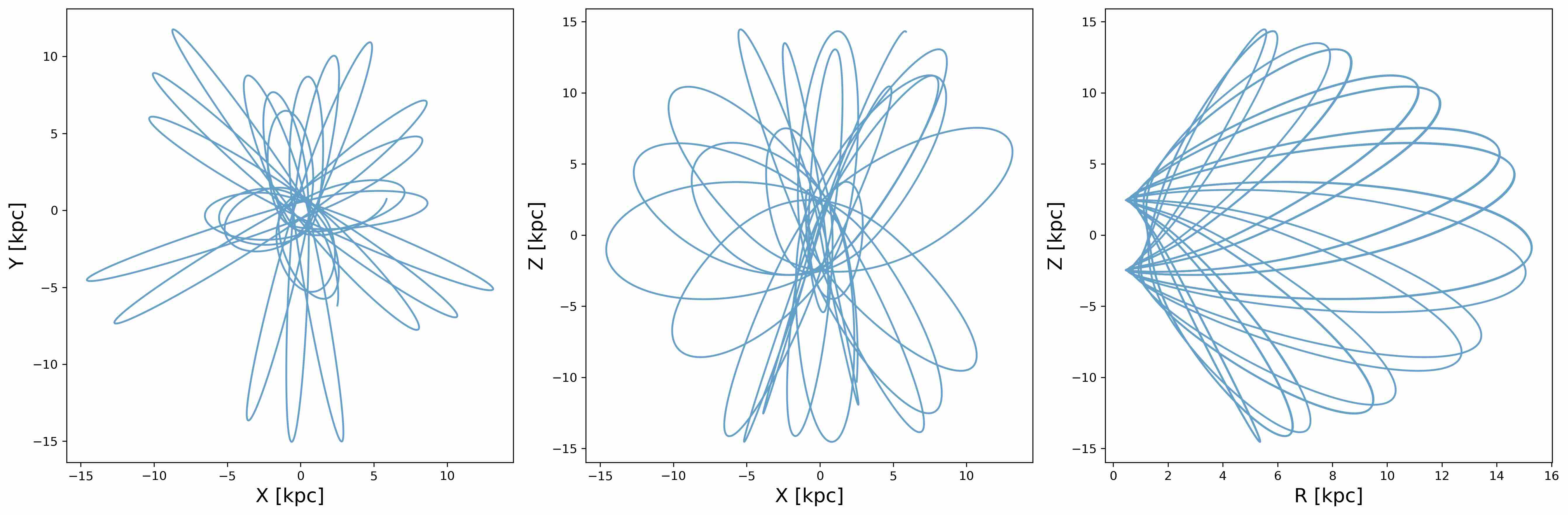}
\caption{Orbit of 2M15203549+0056416 (core candidate 22).}
\label{fig:orbit6}
\end{figure}

\clearpage

\begin{acknowledgments}
This work was supported in part by the National Science Foundation under awards AST-1831412 and AST-2219090, and by the Simons Foundation under award 00533845, as part of the AstroCom NYC program. This work was partially supported by the Sloan Digital Sky Survey’s Faculty And Student Team (FAST) program, funded by the Alfred P. Sloan Foundation. A.A.S., L.N., V.S., and K.C. acknowledge that their work here is part of an SDSS FAST team and is supported, in part, by the National Science Foundation through NSF grant No. AST-2009507. A.A.S., S.P., and A.P.W. also acknowledge insightful discussions during the CCA Nearby Universe Group Meetings. R.G. gratefully acknowledges support from ANID Fondecyt Postdoctoral Grant No. 3230001.

This work has made use of data from the European Space Agency (ESA) mission
{\it Gaia} (\url{https://www.cosmos.esa.int/gaia}), processed by the {\it Gaia}
Data Processing and Analysis Consortium (DPAC,
\url{https://www.cosmos.esa.int/web/gaia/dpac/consortium}). Funding for the DPAC
has been provided by national institutions, in particular the institutions
participating in the {\it Gaia} Multilateral Agreement.

Funding for the Sloan Digital Sky Survey IV has been provided by the Alfred P. Sloan
Foundation, the U.S. Department of Energy Office of Science, and the Participating
Institutions. 

SDSS-IV acknowledges support and resources from the Center for High Performance
Computing  at the University of Utah. The SDSS website is www.sdss4.org.

SDSS-IV is managed by the Astrophysical Research Consortium for the Participating
Institutions of the SDSS Collaboration including the Brazilian Participation Group, the
Carnegie Institution for Science, Carnegie Mellon University, Center for Astrophysics |
Harvard \& Smithsonian, the Chilean Participation Group, the French Participation Group,
Instituto de Astrof\'isica de Canarias, The Johns Hopkins University, Kavli Institute
for the Physics and Mathematics of the Universe (IPMU) / University of Tokyo, the Korean
Participation Group, Lawrence Berkeley National Laboratory, Leibniz Institut f\"ur
Astrophysik Potsdam (AIP),  Max-Planck-Institut f\"ur Astronomie (MPIA Heidelberg),
Max-Planck-Institut f\"ur Astrophysik (MPA Garching), Max-Planck-Institut f\"ur
Extraterrestrische Physik (MPE), National Astronomical Observatories of China, New
Mexico State University, New York University, University of Notre Dame, Observat\'ario
Nacional / MCTI, The Ohio State University, Pennsylvania State University, Shanghai
Astronomical Observatory, United Kingdom Participation Group, Universidad Nacional
Aut\'onoma de M\'exico, University of Arizona, University of Colorado Boulder,
University of Oxford, University of Portsmouth, University of Utah, University of
Virginia, University of Washington, University of Wisconsin, Vanderbilt University, and
Yale University.
This work was supported by a research grant (VIL53081) from VILLUM FONDEN.
\end{acknowledgments}

\bibliographystyle{aasjournal}
\bibliography{bibliography}



\end{document}